\documentclass[a4paper,12pt]{article}

\usepackage{amsmath}
\usepackage[psamsfonts]{amssymb}
\usepackage{rsfs}
\usepackage{bm}

\usepackage{cite}
\usepackage[dvips]{graphicx}
\usepackage{color}

\begin{document} 

\begin{titlepage}

\baselineskip 10pt
\hrule 
\vskip 5pt
\leftline{}
\leftline{Chiba Univ. Preprint
          \hfill   \small \hbox{\bf CHIBA-EP-152}}
\leftline{\hfill   \small \hbox{hep-th/0504088}}
\leftline{\hfill   \small \hbox{May 2005}}
\vskip 5pt
\baselineskip 14pt
\hrule 
\vskip 1.0cm
\centerline{\Large\bf 
Weak gauge-invariance of      
} 
\vskip 0.3cm
\centerline{\Large\bf  
 dimension two condensate
}
\vskip 0.3cm
\centerline{\Large\bf  
in Yang-Mills theory 
}
\vskip 0.3cm
\centerline{\large\bf  
}

\vskip 0.5cm

\centerline{{\bf 
Kei-Ichi Kondo$^{\dagger,{1}}$ 
}}  
\vskip 0.5cm
\centerline{\it
${}^{\dagger}$Department of Physics, Faculty of Science, 
Chiba University, Chiba 263-8522, Japan
}
\vskip 1cm

\begin{abstract}
We give a formal proof that the space-time average of  the vacuum condensate of mass dimension two, i.e.,
the vacuum expectation value of the squared potential $\mathscr{A}_\mu^2$,  is gauge invariant  in the weak sense that it is independent of the gauge-fixing condition adopted in quantizing the Yang-Mills theory.
This is shown at least for the small deformation from the generalized Lorentz and the modified Maximal Abelian gauge in the naive continuum formulation neglecting Gribov copies. 
This suggests that the numerical value of the condensate could be the same no matter what gauge-fixing conditions for choosing the representative from the gauge orbit are adopted  to measure it. 
Finally, we discuss how this argument should be modified when the Gribov copies exist. 

\end{abstract}


\hrule  
\vskip 0.1cm
${}^1$ 
  E-mail:  {\tt kondok@faculty.chiba-u.jp}

\par 
\par\noindent


\vskip 0.5cm

\newpage
\pagenumbering{roman}




\end{titlepage}


\pagenumbering{arabic}

\baselineskip 14pt
\section{Introduction}

In the gauge theory, only the gauge invariant quantities are observable, since gauge non-invariant quantities change their values by the gauge transformation and can not take definite values.  
The dimension-three quark-antiquark bilinear operator 
$\bar{\psi}(x) \psi(x)$ and the dimension-four gluon operator 
${\rm tr} \left( \mathscr{F}_{\mu\nu}(x)\mathscr{F}^{\mu\nu}(x)\right)$ 
are typical gauge-invariant operators  in QCD.  They are local operators defined on a spacetime point. 

We wish to discuss how the gauge noninvariant operator may be gauge invariant. 
If we do not restrict to the local operator, we can enlarge the candidates of the gauge invariant operator.
In this paper, we focus on composite operators of mass dimension two, since they were taken up by many papers from the analytical and numerical viewpoints, see e.g., \cite{GH86,LS88,Boucaudetal00,GSZ01,Schaden99,Dudaletal04,ABB04,FN05,NW05,CN05}. 
For example, we know that the squared potential 
${\rm tr} \left(  \mathscr{A}_\mu(x)^2 \right)$
of Yang-Mills theory  
is gauge variant.  
However, the minimized squared potential
$\min_{\omega} \int d^4x {\rm tr} \left(  \mathscr{A}_\mu^\omega(x)^2 \right)$ with respect to the gauge transformation $\omega$ in Euclidean space is clearly gauge invariant \cite{GSZ01}, since ${\rm tr} \left(  \mathscr{A}_\mu(x)^2 \right)$ can not be negative and has a non-negative minimum and hence $\min_{\omega} \int d^4x {\rm tr} \left(  \mathscr{A}_\mu^\omega(x)^2 \right)$ no longer change the value under the gauge transformation. 
Thus, the composite operator of mass dimension two can be gauge invariant by this procedure in classical gauge theory.

In  quantum theory, however, we can not measure the value of the operator itself and can measure only the vacuum expectation value of the operator. 
In this sense, the gauge invariance of the operator stated in the above ({\it strong gauge invariance in the operator level}) is too  strong.  
If the gauge theory is written in terms of the gauge potential, we must fix the gauge in quantizing the gauge theory and the Becchi-Rouet-Stora-Tyutin (BRST) symmetry in quantum gauge theory plays a role of the gauge symmetry of classical gauge theory. 
Therefore, we can check whether the (elementary or composite) operator is gauge invariant or not, respectively, by examining whether or not the vacuum expectation value of the operator keeps the value under deforming gauge fixing condition. 
It is the {\it weak gauge invariance} ({\it the gauge-fixing-condition independence}) that we wish to discuss in this paper. 

A first step in this direction shown in \cite{Kondo01} is a fact that the spacetime average of the mixed gluon--ghost condensate of mass dimension two,
\begin{align}
  \left< \mathscr{O}_{K} \right> = \mathcal{V}^{-1} \int_{\mathcal{V}} d^4x \left<  {\rm tr}_{G/H} \left( \frac{1}{2} \mathscr{A}_\mu(x) \cdot \mathscr{A}^\mu(x) - \lambda i \mathscr{C}(x) \cdot \bar{\mathscr{C}}(x) \right)  \right> , 
\end{align}
is BRST invariant in the generalized Lorentz gauge and the modified Maximal Abelian gauge, although  
the operator 
${\rm tr} \left( \frac{1}{2} \mathscr{A}_\mu(x)^2 \right)$
is gauge variant.  
This claim is based on the on-shell BRST invariance of the operator $\mathscr{O}_{K}$. 
Here it is essential to take both the spacetime average and the vacuum expectation value.

Recently, Slavnov \cite{Slavnov04} has shown that the spacetime average of the dimension two condensate,
\begin{align}
  \mathcal{V}^{-1} \int_{\mathcal{V}} d^4x \left<  {\rm tr} \left( \frac{1}{2} \mathscr{A}_\mu(x)^2 \right) \right> 
  := \left<\left< {\rm tr} \left( \frac{1}{2} \mathscr{A}_\mu^2 \right) \right>\right>  ,
\end{align}
is {\it gauge invariant} by using a noncommutative field theory technique under three hypothesis. 
Subsequently, he has shown \cite{Slavnov05} that the  averaged condensate 
$\left<\left<    \frac{1}{2} A_\mu^2   \right>\right>$
in the Abelian gauge theory
is  {\it gauge independent}, i.e., {\it gauge-fixing-parameter independent}  without using the noncommutative technique and in the standard commutative formalism, i.e., 
the averaged condensate does not change at least in the Abelian gauge theory when the gauge-fixing-parameter  is changed. 

In view of these, we expect that the 
${\rm tr} \left( \frac{1}{2} \mathscr{A}_\mu(x)^2 \right)$
can become gauge invariant only after taking both the spacetime average and the vacuum expectation value, i.e.,
$\mathcal{V}^{-1} \int_{\mathcal{V}} d^4x \left<  {\rm tr} \left( \frac{1}{2} \mathscr{A}_\mu(x)^2 \right) \right> $. 
Note that either the spacetime average, 
$\mathcal{V}^{-1} \int_{\mathcal{V}} d^4x  {\rm tr} \left( \frac{1}{2} \mathscr{A}_\mu(x)^2 \right)$,
or the vacuum expectation value,
$\left<  {\rm tr} \left( \frac{1}{2} \mathscr{A}_\mu(x)^2 \right) \right>$, 
is not expected to be gauge invariant. 

What we wish to discuss in this paper is the {\it gauge-fixing-condition independence} of the dimension two condensate, rather than the gauge-fixing-parameter independence, in the non-Abelian gauge theory in the standard commutative Yang-Mills theory formalism. 

In this paper, we give a formal proof that the space-time average of a vacuum condensate of mass dimension two  $\left< \mathscr{O}_{K} \right>$
is gauge invariant  in the weak sense that it is independent of the gauge-fixing condition adopted in quantizing the Yang-Mills theory, at least for the small deformation from the generalized Lorentz and the modified Maximal Abelian gauge when we neglect Gribov copies.  
In the final section, we examine if the statement holds by restricting the functional integral to the Gribov region or the fundamental modular region, when the Gribov copies exist.



\section{Gauge-fixing-condition independence}
\setcounter{equation}{0}

For our purpose, it is convenient to adopt the quantization method based on the functional integration. 
The vacuum expectation value of the operator $\mathscr{O}$ is defined by the functional integral:
\begin{align}
      \left<0| \mathscr{O} |0\right> 
=   Z^{-1}_{YM} \int d\mu_{YM} \  e^{iS_{YM}+iS_{GF+FP}}  \mathscr{O}   ,
\end{align}
with the integration measure $d\mu_{YM}$ given by 
\begin{align}
  d\mu_{YM} 
  = [d\mathscr{A}_\mu] [d\mathscr{N}] [d\mathscr{C}] [d\bar{\mathscr{C}}] ,
\end{align}
where $\mathscr{A}_\mu, \mathscr{N}, \mathscr{C}, \bar{\mathscr{C}}$ are respectively the gauge field, the Nakanishi--Lautrup (NL) auxiliary field, the Faddeev--Popov (FP) ghost field and antighost field. 
$Z_{YM}$ guarantees the normalization $\left<0| 1 |0\right>=1$.
Here, $S_{YM}$ is the gauge-invariant Yang-Mills action, while $S_{GF+FP}$ is the sum of the gauge-fixing (GF) and the associated FP ghost terms.  

The BRST transformation $\bm{\delta}_{B}$ is defined so that $S_{YM}$ is BRST invariant. Then $S_{GF+FP}$ is also BRST invariant, since it is written in the BRST-exact form using the nilpotent BRST transformation, i.e., $\bm{\delta}_{B}^2=0$. 
Note that the measure $[d\mathscr{A}_\mu]$ is gauge invariant and  $d\mu_{YM}$ is BRST invariant. 

For the gauge-fixing condition $F[\mathscr{A}]=0$, 
we consider the BRST-exact $S_{GF+FP}$ term which has the form,
\begin{align}
  S_{GF+FP} = \int d^4x \ i^{-1} \bm{\delta}_{B} 
  \left\{ \bar{\mathscr{C}}(x) \cdot \left( F[\mathscr{A}(x)] + \frac{\lambda}{2} G[\mathscr{N}(x),\mathscr{C}(x),\bar{\mathscr{C}}(x)]  
   \right)  \right\} ,
\end{align}
where $\lambda$ is the gauge-fixing parameter. 
In the usual covariant Lorentz gauge, we see 
\begin{align}
  F[\mathscr{A}] = \partial^\mu \mathscr{A}_\mu,  \quad
  G[\mathscr{N},\mathscr{C},\bar{\mathscr{C}}] = \mathscr{N} .
\end{align}

Now we introduce an operation $\delta_{F}$ of changing the gauge fixing condition 
$F[\mathscr{A}]=0$ ($\lambda$-independent part) infinitesimally.
We apply this operation to the vacuum expectation value 
$\left<0| \mathscr{O} |0\right>$ 
of an elementary or composite operator $\mathscr{O}$.  
Then we obtain
\begin{align}
&  \delta_{F} \left<0| \mathscr{O} |0\right> 
  \nonumber\\
  =& \left<0\left| \mathscr{O} \int d^4y  \bm{\delta}_{B} \{\bar{\mathscr{C}}(y)  \cdot \delta_{F} F[\mathscr{A}(y)] \} \right|0\right> 
  - \left<0| \mathscr{O} |0\right> \left<0\left| \int d^4y  \bm{\delta}_{B} \{ \bar{\mathscr{C}}(y)  \cdot \delta_{F} F[\mathscr{A}(y)] \} \right|0\right> 
  \nonumber\\
  =& \left<0\left| \mathscr{O} \int d^4y  \bm{\delta}_{B} \{\bar{\mathscr{C}}(y)  \cdot \delta_{F} F[\mathscr{A}(y)] \} \right|0\right> ,
\end{align}
since the BRST transformation is generated by the BRST charge $Q_B$ according to 
\begin{align}
  \bm{\delta}_{B}(*) = [ iQ_B, *]_{\mp} ,  
  \label{QB}
\end{align}
and the vacuum is annihilated by the BRST charge (Kugo--Ojima subsidiary condition):
\begin{align}
  Q_B |0 \rangle = 0 .
  \label{subc}
\end{align}

Taking into account an identity
\begin{align}
 \left<0\left| \bm{\delta}_{B} \left( \mathscr{O} \int d^4y   \{\bar{\mathscr{C}}(y)  \cdot \delta_{F} F[\mathscr{A}(y)] \} \right) \right|0\right> = 0 , 
\end{align}
following from (\ref{QB}) and (\ref{subc}), 
we obtain a basic relationship,
\begin{align}
  \delta_{F} \left<0| \mathscr{O} |0\right> 
  =  \mp \left<0\left| \bm{\delta}_{B}\mathscr{O} \int d^4y  \{ \bar{\mathscr{C}}(y)  \cdot \delta_{F} F[\mathscr{A}(y)] \} \right|0\right> ,
\label{change}
\end{align}
where the $-(+)$ in the right-hand side should be understood for an operator $\mathscr{O}$ with zero and even (odd) ghost number.

The eq.(\ref{change}) implies that, {\it if an operator $\mathscr{O}$ is BRST-invariant $\bm{\delta}_{B}\mathscr{O}=0$,  the vacuum expectation value of the  operator $\mathscr{O}$ does not depend on the gauge-fixing condition (procedure).}  
Hence, the BRST-invariant operator is gauge invariant in this restricted sense. 
This corresponds to the opposite of the statement in the operator level that the gauge-invariant operator is BRST-invariant. 
In general, BRST-invariant operator is not necessarily gauge invariant.   
In our approach, it is essential to consider the vacuum expectation value as a criterion of gauge invariance of the operator (weak gauge invariance), which is weaker than the gauge invariance in the operator level (strong gauge invariance).
In this paper, we check the gauge invariance of the operator in the weak sense, by first breaking the gauge through the gauge fixing procedure  and then seeing the stability of the vacuum expectation value under the arbitrary deformation of the gauge fixing.

\subsection{Generalized Lorentz gauge}

In what follows, we focus on the mixed gluon-ghost operator of mass dimension two defined by \cite{Kondo01}
\begin{align}
  \mathscr{O}_{K} := \mathcal{V}^{-1} \int_{\mathcal{V}} d^4x \left( \frac{1}{2} \mathscr{A}_\mu \cdot \mathscr{A}^\mu - \lambda i \mathscr{C} \cdot \bar{\mathscr{C}}  \right) , 
\end{align}
in the generalized Lorentz gauge \cite{Kondo01},
\begin{align}
  S_{GF+FP}^{} =  \int d^4x \ i \bm{\delta}_{B} \bar{\bm{\delta}}_{B} \left( \frac{1}{2} \mathscr{A}_\mu \cdot \mathscr{A}^\mu - \frac{\lambda}{2} i \mathscr{C} \cdot \bar{\mathscr{C}}  \right) , 
\end{align}
with the anti-BRST transformation $\bar{\bm{\delta}}_{B}$.
The generalized Lorentz gauge is equivalent to take
\begin{align}
  F[\mathscr{A}] = \partial^\mu \mathscr{A}_\mu,  \quad
  G[\mathscr{N},\mathscr{C},\bar{\mathscr{C}}] = \mathscr{N} -i \frac{g}{2} (\bar{\mathscr{C}} \times \mathscr{C}) . 
\end{align}
At $\lambda=0$, the generalized Lorentz gauge coincides with the usual Lorentz gauge, i.e., the Landau gauge. 

It has been shown \cite{Kondo01} that the operator $\mathscr{O}_{K}$ is on-shell BRST invariant, but it is not (off-shell) BRST invariant, $ \bm{\delta}_{B}\mathscr{O}_{K} \not=0$. 
The on-shell BRST transformation is defined for $\mathscr{A}_\mu,\mathscr{C}, \bar{\mathscr{C}}$ and is obtained by eliminating the NL field $\mathscr{N}$ from the off-shell BRST transformation through the equation of motion for $\mathscr{N}$. 
The GF+FP term can be rewritten in terms only of $\mathscr{A}_\mu,\mathscr{C}, \bar{\mathscr{C}}$ and the resulting term is checked to be on-shell BRST invariant, but it is not written in the on-shell BRST-exact form. 
Moreover, the on-shell BRST transformation is not nilpotent, unless the equation of motion for the ghost is used.  
In this paper, we do not use the on-shell BRST transformation. 

The key identity we use in the following is 
\begin{align}
  \bm{\delta}_{B} \mathscr{O}_{K} 
  = - \mathcal{V}^{-1} \int_{\mathcal{V}} d^4x \ \mathscr{C}(x) \cdot 
  \frac{\delta S_{GF+FP}}{\delta \mathscr{N}(x) } .
\label{form}
\end{align}
In fact, the BRST transform of the operator reads
\begin{align}
  \bm{\delta}_{B}\mathscr{O}_{K} 
  =& \mathcal{V}^{-1} \int_{\mathcal{V}} d^4x \left(  \mathscr{A}_\mu \cdot \bm{\delta}_{B}\mathscr{A}^\mu 
  - \lambda i \bm{\delta}_{B}\mathscr{C} \cdot \bar{\mathscr{C}}   
  + \lambda i \mathscr{C} \cdot \bm{\delta}_{B}\bar{\mathscr{C}}  \right)
\nonumber\\ 
  =& \mathcal{V}^{-1} \int_{\mathcal{V}} d^4x \left[  \mathscr{A}_\mu \cdot D^\mu[\mathscr{A}] \mathscr{C}  
  - \lambda i \frac{-g}{2} (\mathscr{C} \times \mathscr{C}) \cdot \bar{\mathscr{C}}   
  + \lambda i \mathscr{C} \cdot i\mathscr{N}   \right]
\nonumber\\ 
  =& \mathcal{V}^{-1} \int_{\mathcal{V}} d^4x \left[  \mathscr{A}_\mu \cdot \partial^\mu  \mathscr{C}  
  + \lambda i \frac{g}{2}  \mathscr{C}  \cdot (\mathscr{C} \times \bar{\mathscr{C}})  
  - \lambda   \mathscr{C} \cdot  \mathscr{N}   \right]
\nonumber\\ 
  =& \mathcal{V}^{-1} \int_{\mathcal{V}} d^4x \  \mathscr{C} \cdot \left[ - \partial^\mu  \mathscr{A}_\mu  
  + \lambda i \frac{g}{2} (\mathscr{C} \times \bar{\mathscr{C}})  
 - \lambda  \mathscr{N}   \right] .
\end{align}
If this is combined  with the fact,
\begin{align}
  \frac{\delta S_{GF+FP}}{\delta \mathscr{N}(x)} 
  =   \partial^\mu \mathscr{A}_\mu(x) - \lambda i \frac{g}{2} (\mathscr{C} \times \bar{\mathscr{C}})(x) + \lambda \mathscr{N}(x)  ,
\end{align}
the desired equality (\ref{form}) is obtained.

Therefore, by substituting (\ref{form}) into (\ref{change}), we obtain
\begin{align}
 &  \delta_{F} \left<0| \mathscr{O}_{K} |0\right> 
 \nonumber\\
  =&  \mathcal{V}^{-1} \int_{\mathcal{V}} d^4x
   \left<0\left| \frac{\delta}{\delta \mathscr{N}^A(x) } \left[ \mathscr{C}^A(x) S_{GF+FP}  \int d^4y   \bar{\mathscr{C}}(y)  \cdot \delta_{F} F[\mathscr{A}(y)]   \right] \right|0\right> 
   \nonumber\\
  =&  \mathcal{V}^{-1} \int_{\mathcal{V}} d^4x
   Z^{-1}_{YM} \int d\mu_{YM} \  e^{iS_{YM}+iS_{GF+FP}}  \frac{\delta}{\delta \mathscr{N}^A(x) } \left[ \mathscr{C}^A(x) S_{GF+FP} \int d^4y   \bar{\mathscr{C}}(y)  \cdot \delta_{F} F[\mathscr{A}(y)]  \right]   
   \nonumber\\
  =&  \mathcal{V}^{-1} \int_{\mathcal{V}} d^4x
   Z^{-1}_{YM} \int d\mu_{YM} \  \frac{\delta}{\delta \mathscr{N}^A(x) } \left[ -i \mathscr{C}^A(x) e^{iS_{YM}+iS_{GF+FP}}  \int d^4y   \bar{\mathscr{C}}(y)  \cdot \delta_{F} F[\mathscr{A}(y)]  \right]   .
   \label{identity}
\end{align}
Note that the right-hand side is zero.%
\footnote{
The integration of the differentiation is identically zero,
$
 \int \mathcal{D}\mathscr{N} {\delta \over \delta \mathscr{N}}f(\mathscr{N})=0  
$.
This follows only from the shift invariance of the measure, i.e.,
$\mathcal{D}(\varphi+a)=\mathcal{D}\varphi$ for arbitrary $a$, since for an arbitrary functional $f(\varphi)$ of $\varphi$ it implies that 
$ \int \mathcal{D}\varphi f(\varphi)=\int \mathcal{D}(\varphi+a) f(\varphi+a) 
=\int \mathcal{D}\varphi f(\varphi) + a \int \mathcal{D}\varphi {\delta \over \delta \varphi}f(\varphi)+ \cdots $
should hold for arbitrary $a$, 
and hence 
$
 \int \mathcal{D}\varphi {\delta \over \delta \varphi}f(\varphi)=0  
$
is obtained. 
This also follows if the boundary values $f(\varphi_{max})$, $f(\varphi_{min})$ of the functional  vanish, since 
$
 \int \mathcal{D}\varphi {\delta \over \delta \varphi}f(\varphi)
 = f(\varphi_{max})-f(\varphi_{min}) . 
$

}  
Therefore, we conclude 
\begin{align}
   \delta_{F} \left<0| \mathscr{O}_{K} |0\right>  = 0 ,
\end{align}
which means that the spacetime average of the vacuum expectation value of mass dimension two: 
\begin{align}
  \left< \mathscr{O}_{K} \right> = \mathcal{V}^{-1} \int_{\mathcal{V}} d^4x \left<  \frac{1}{2} \mathscr{A}_\mu(x) \cdot \mathscr{A}^\mu(x) - \lambda i \mathscr{C}(x) \cdot \bar{\mathscr{C}}(x)   \right> , 
\end{align}
is unchanged even if we adopt the gauge-fixing condition which is slightly deformed from the original one. 
In particular for $\lambda=0$, starting from the Landau gauge in the usual Lorentz gauge fixing, we see the invariance of the vacuum condensate
\begin{align}
  \left< \mathscr{O}_{K} \right> = \mathcal{V}^{-1} \int_{\mathcal{V}} d^4x \left<  \frac{1}{2} \mathscr{A}_\mu(x) \cdot \mathscr{A}^\mu(x)     \right> . 
\end{align}
for the deformation of the Landau gauge-fixing condition 
$\partial^\mu \mathscr{A}_\mu=0$.

\begin{figure}[htbp]
\begin{center}
\includegraphics[height=5cm]{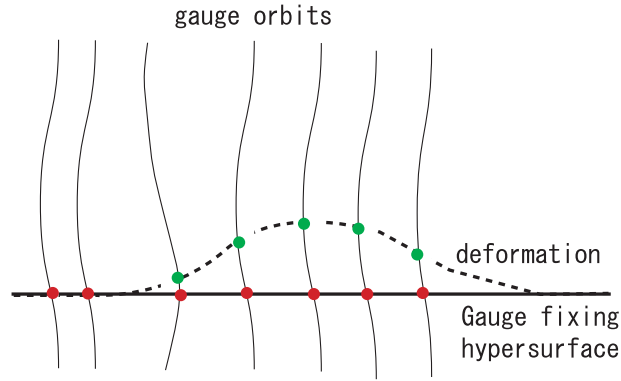}
\caption{\small 
Gauge orbits and the gauge-fixing hypersuface (solid line) specified by the gauge-fixing condition and its infinitesimal deformation (broken line).}
\label{fig:orbit-GF}
\end{center}
\end{figure}

The gauge fixing is performed by choosing a representative from a gauge orbit (gauge equivalent configurations of the given gauge potential) as an intersecting point of a gauge orbit with a gauge-fixing hypersurface specified by the gauge-fixing condition.  This is clear for $\lambda=0$ case. 
Any gauge-fixing hypersurface is obtained by repeating the infinitesimal continuous change from an initial gauge-fixing hypersurface.  Therefore, the gauge-fixing independence of $\left< \mathscr{O} \right>$ implies that the value of $\left< \mathscr{O} \right>$ has a unique value and  does not change no matter how we choose one representative from each gauge orbit.  
In this sense, the gauge-fixing independence leads to the gauge invariance.  
See Fig.~\ref{fig:orbit-GF}.


\subsection{Modified Maximal Abelian gauge}
\par

For the gauge group $G=SU(N)$, we consider the Cartan decomposition of $\mathscr{A}$ into diagonal (maximal torus group $H=U(1)^{N-1}$) and off-diagonal ($G/H$) pieces,
\begin{equation}
 \mathscr{A}(x) = \mathscr{A}^A(x) T^A
 = a^i(x) H^i + A^a(x) T^a , \quad 
 (A=1, \cdots, N^2-1) ,
\end{equation}
where $i=1, 2, \cdots, N-1$. 
In particular, for $G=SU(3)$, the generators are given by the Gell-Mann matrices:
\begin{equation}
 H^1={\lambda^3 \over 2}, \quad H^2={\lambda^8 \over 2}, \quad
 T^a = {\lambda^a \over 2}   \ (a=1,2,4,5,6,7) .
\end{equation}
The  Maximal Abelian (MA) gauge-fixing condition plays a role of the partial gauge fixing $G \rightarrow H$ and is given by
\begin{equation}
 F^a[a,A] := (D_\mu[a] A^\mu)^a
 = \partial_\mu A^\mu{}^a(x) + g f^{aib}a_\mu^i(x) A^\mu{}^b(x)  =0 .
\end{equation}
The naive MA gauge is given by
\begin{eqnarray}
  S_{GF+FP} = - \int d^4x i \bm{\delta}_{B} 
\left[ \bar C^a \left( D^\mu[a]A_\mu + {\alpha \over 2} N \right)^a
 \right] .
\label{MA}
\end{eqnarray}
In order to fix the residual Abelian gauge group $H$, we add an additional $GF+FP$ term for the diagonal part, e.g.,
\begin{align}
- \int d^4x i \bm{\delta}_{B} \left[ \bar C^i \left( \partial^\mu a_\mu + {\beta \over 2} N \right)^i   \right] .
\end{align}

We consider the operator
\begin{align}
  \mathscr{O}_{K}' := \mathcal{V}^{-1} \int_{\mathcal{V}} d^4x{\rm tr}_{G/H}  \left( \frac{1}{2} \mathscr{A}_\mu \mathscr{A}^\mu - \alpha i \mathscr{C} \bar{\mathscr{C}}  \right) , 
\end{align}
in the modified MA gauge defined by \cite{KondoII}
\begin{eqnarray}
  S_{GF+FP}' 
  := \int d^4x \ i \bm{\delta}_{B} \bar{\bm{\delta}}_{B}
  {\rm tr}_{G/H} \left[ {1 \over 2} \mathscr{A}_\mu \mathscr{A}^\mu
  -{\alpha \over 2}i \mathscr{C} \bar{\mathscr{C}} \right] .
  \label{mMA}
\end{eqnarray}
Note that (\ref{mMA}) is obtained  from (\ref{MA}) by adding the
ghost self-interaction terms and by adjusting the parameter for the ghost self-interaction term,
\begin{align}
  F^a[\mathscr{A}] = (D^\mu[a] A_\mu)^a ,  \quad
  G^a[\mathscr{N},\mathscr{C},\bar{\mathscr{C}}] = N^a
- i g  f^{abi} \bar C^b C^i
- i {g \over 2} f^{abc}  C^b  \bar C^c , 
\end{align}
since
\begin{eqnarray}
 &&  \bar{\bm{\delta}}_{B} \left[ {1 \over 2} A_\mu^a A^\mu{}^a
  -{\alpha \over 2} i C^a \bar C^a \right]
\nonumber\\
&=& - \bar C^a \left( D^\mu[a] A_\mu  +   {\alpha \over 2}N \right)^a 
+ i {\alpha \over 2} f^{abi} \bar C^a \bar C^b C^i
+ i {\alpha \over 4} f^{abc}  C^a \bar C^b \bar C^c  .
\nonumber\\
&=& - \bar C^a \left[ (D^\mu[a] A_\mu)^a  
+   {\alpha \over 2} \left( N^a
- i g  f^{abi} \bar C^b C^i
- i {g \over 2} f^{abc}  C^b  \bar C^c \right) 
 \right] ,
\end{eqnarray}
where the structure constants $f^{ABC}$ are completely antisymmetric in the indices and  we have used a fact 
$f^{aij}=0$ ($T^i$ and $T^j$ commute).  

It is easy to check that the similar identity holds in the modified MA gauge,
\begin{align}
  \bm{\delta}_{B} \mathscr{O}_{K}' 
  =& - \mathcal{V}^{-1} \int_{\mathcal{V}} d^4x \ C^a(x)  
  \frac{\delta S_{GF+FP}'}{\delta N^a(x) } ,
\label{form2}
\end{align}
where
\begin{align}
 \bm{\delta}_{B} \mathscr{O}_{K}' 
  =  - \mathcal{V}^{-1} \int_{\mathcal{V}} d^4x \  C^a   \left[   (D^\mu  A_\mu)^a  
  -  \alpha g  f^{abi} i  C^i \bar C^b
  + \alpha   \frac{g}{2} f^{abc} i C^b \bar{C}^c   
 + \alpha  N^a   \right] .
\end{align}
In the same way as in the Lorentz gauge, therefore, we conclude 
\begin{align}
   \delta_{F} \left<0| \mathscr{O}_{K}' |0\right>  = 0 ,
\end{align}
which means that the spacetime average of the vacuum expectation value of mass dimension two: 
\begin{align}
  \left< \mathscr{O}_{K}' \right> = \mathcal{V}^{-1} \int_{\mathcal{V}} d^4x \left<  \frac{1}{2} A_\mu^a(x) A^\mu{}^a(x)
  -{\alpha \over 2} i C^a(x) \bar C^a(x)   \right> , 
\end{align}
is unchanged even if we adopt the gauge-fixing condition which is slightly deformed from the original one. 
In particular for $\alpha=0$, starting from the Landau gauge in the naive MA gauge, we see the invariance of the vacuum condensate
\begin{align}
  \left< \mathscr{O}_{K}' \right> = \mathcal{V}^{-1} \int_{\mathcal{V}} d^4x \left<  \frac{1}{2} A_\mu^a(x) A^\mu{}^a(x)     \right> , 
\end{align}
for the deformation of the gauge-fixing condition.

\section{Conclusion and discussion}

In this paper, we have given a formal proof that the space-time average of a vacuum condensate of mass dimension two  $\left< \mathscr{O}_{K} \right>$
is gauge invariant  in the weak sense that it is independent of the gauge-fixing condition adopted in quantizing the Yang-Mills theory, at least for the small deformation from the generalized Lorentz and the modified MA gauge  
in the naive continuum formulation without Gribov copies.

This suggests that the numerical value of the condensate must be the same no matter what gauge-fixing conditions for choosing the representative from the gauge orbit are adopted  to measure it. We discuss this point for a while.

The Lorentz gauge and the MA gauge can be interpolated by introducing a parameter $\xi$ \cite{KondoI} for the off-diagonal gauge-fixing condition, 
\begin{align}
  F^a_{\xi}  := \partial^\mu A_\mu^a(x) + \xi g f^{aib}  a^\mu{}^i(x) A_\mu^b(x)  =0 ,
\end{align}
and by taking the same Landau gauge condition for the diagonal part.
\begin{align}
  F^i_{\xi}   
 := \partial^\mu a_\mu^i(x)   =0 ,
\end{align}
where $\xi=0$ is the Lorentz gauge and $\xi=1$ is the MA gauge. Then
$\delta_{F}F^A =\xi g f^{Aib}  a^\mu{}^i A_\mu^b$. 
Therefore, for small change of $\xi$ from 0 and 1, the vacuum condensation of mass dimension two must not change the value, i.e.,  
$\delta_{F} \left<0| \mathscr{O}_{K} |0\right>  =\mathcal{O}(\xi^2).$

Is it possible to show the above independence also for a finite deformation of the gauge-fixing condition?  
This is trivial for a finite deformation, if it is obtained by performing the infinitesimal transformation successively, as far as the obstruction for the deformation does not exist.  
In order to treat the finite deformation,  
we denote the vacuum expectation value calculated under the gauge-fixing condition $F=0$ by 
$\left<0| \mathscr{O}_{K} |0\right>_{F}$. 
If we consider the change of gauge-fixing condition $F$ by $\delta_{F}F$, the relationship is obtained:
\begin{align}
   \left<0| \mathscr{O}_{K} |0\right>_{F+\delta_{F}F} 
   = \frac{\left<0| \mathscr{O}_{K} e^{i\delta_{F}S_{GF+FP}}|0\right>_{F} }{\left<0| e^{i\delta_{F}S_{GF+FP}} |0\right>_{F} } .
\end{align}
The denominator reads
\begin{align}
 \left<0|  e^{i\delta_{F}S_{GF+FP}}|0\right>_{F} 
 =& 1 + \sum_{n=1}^{\infty} \frac{1}{n!} \left<0|  (i\delta_{F}S_{GF+FP})^n  |0\right>_{F}
 \nonumber\\
 =& 1 + \sum_{n=1}^{\infty} \frac{1}{n!} \left<0| \bm{\delta}_{B} [(*)\bm{\delta}_{B}(*) \cdots \bm{\delta}_{B}(*) ]  |0\right>_{F}
 = 1 ,
\end{align}
where we defined 
$*:=\int d^4x  \{\bar{\mathscr{C}}(x) \cdot \delta_{F} F[\mathscr{A}(x)]$
and have used the nilpotency of the BRST transformation. 
On the other hand, the numerator reads
\begin{align}
 \left<0| \mathscr{O}_{K} |0\right>_{F+\delta_{F}F}
 =& \left<0| \mathscr{O}_{K} e^{i\delta_{F}S_{GF+FP}}|0\right>_{F} 
 \nonumber\\
 =& \left<0| \mathscr{O}_{K} |0\right>_{F} + \sum_{n=1}^{\infty} \frac{1}{n!} \left<0| \mathscr{O}_{K} (i\delta_{F}S_{GF+FP})^n  |0\right>_{F}
 \nonumber\\
 =& \left<0| \mathscr{O}_{K} |0\right>_{F} + \sum_{n=1}^{\infty} \frac{1}{n!} \left<0| \mathscr{O}_{K} \bm{\delta}_{B} [(*) \underbrace{\bm{\delta}_{B}(*) \cdots \bm{\delta}_{B}(*) ]}_{n-1}  |0\right>_{F}
 \nonumber\\
 =& \left<0| \mathscr{O}_{K} |0\right>_{F} - \sum_{n=1}^{\infty} \frac{1}{n!} \left<0| \bm{\delta}_{B}\mathscr{O}_{K}  [(*) \underbrace{\bm{\delta}_{B}(*) \cdots \bm{\delta}_{B}(*) ]}_{n-1}  |0\right>_{F} ,
\end{align}
where we have used the nilpotency of the BRST transformation in the third equality. 
The term $*$  does not include the the NL field $\mathscr{N}$.  This is also the case for the $n=1$ piece of this expansion. 
In the $n=2$ piece, however, the NL field $\mathscr{N}$ appears from $\bm{\delta}_{B}\bar{\mathscr{C}}$ in $\bm{\delta}_{B}(*)$. This invalidates the argument given in this paper. 
Therefore, our proof is too naive to extend the independence proof to a finite deformation beyond the infinitesimal deformation of the gauge-fixing condition.


In the above, we implicitly assumed that the gauge-fixing hypersurface intersects the gauge orbit only once.   
It is known, however, that this is not necessarily achieved by the usual gauge-fixing condition, e.g., Landau gauge.  
In fact, there are many intersection points called Gribov copies in a gauge orbit with the given gauge-fixing hypersurface. 
This difficulty is known as the Gribov problem. 
For the generalized Lorentz gauge $\lambda\not=0$, however, the GF+FP term includes the quartic ghost self-interaction term and hence the Gribov problem is not manifest and could be circumvented. 

For the Landau gauge $\lambda=0$, the Gribov copies are shown to exist. 
A simple way (a proposal due to Gribov) to avoid the Gribov copies is to restrict the functional integral into the interior $\Omega$ of the Gribov horizon where the Faddeev--Popov determinant keeps the positive value. 
In the Landau gauge $\lambda=0$ for $F[\mathscr{A}] :=\partial^\mu \mathscr{A}_\mu$, the ghost and anti-ghost field can be integrated out to obtain
\begin{align}
 &  \delta_{F} \left<0| \mathscr{O}_{K} |0\right> 
 \nonumber\\
  =&  \mathcal{V}^{-1} \int_{\mathcal{V}} d^4x
   Z^{-1}_{YM} \int_\Omega [d\mathscr{A}_\mu] \int[d\mathscr{N}]
    \  \frac{\delta}{\delta \mathscr{N}^A(x) } 
    \Big[ \int d^4y \ i^{-1} \delta_{F} F^B[\mathscr{A}(y)] e^{iS_{YM}}
 \nonumber\\& \quad\quad\quad\quad 
  \times \int [d\mathscr{C}] [d\bar{\mathscr{C}}]   \mathscr{C}^A(x) \bar{\mathscr{C}}^B(y) e^{iS_{GF+FP}}      \Big]    
 \nonumber\\
  =&  \mathcal{V}^{-1} \int_{\mathcal{V}} d^4x
   Z^{-1}_{YM} \int_\Omega [d\mathscr{A}_\mu] \int[d\mathscr{N}]
    \  \frac{\delta}{\delta \mathscr{N}^A(x) } 
    \Big[ \int d^4y \ i^{-1} \delta_{F} F^B[\mathscr{A}(y)] e^{iS_{YM}}
 \nonumber\\& \quad\quad\quad\quad 
  \times e^{i\int d^4x \mathscr{N} \cdot F[\mathscr{A}]}  
  [-\partial \cdot D[\mathscr{A}]]^{-1}_{AB}(x,y)
   \det  [-\partial \cdot D[\mathscr{A}]]
      \Big]   ,
\end{align}
where we have used the identity (\ref{identity}) and the ghost and anti-ghost field are integrated out.  
The right-hand side is zero, since the integration of the total derivative with respect to $\mathscr{N}$ vanishes. 
Within the Gribov proposal, therefore, the gauge-fixing independence is shown for arbitrary gauge-fixing hypersurface which is deformed from the hypersurface  $\partial^\mu \mathscr{A}_\mu=0$ inside the Gribov horizon of the Landau gauge fixing. 

However, this reasoning is clearly insufficient, since it is proved that there is no Gribov copies only in a subset of the Gribov region, called the  fundamental modular region $H$. 
In the rigorous sense, therefore, we must consider the functional integration restricted to the fundamental modular region $H$.  
In fact, in gauge-fixed lattice simulations, the fundamental modular region is   selected and used to measure the averaged condensate, 
\noindent$\displaystyle \mathcal{V}^{-1}\int_\mathcal{V} d^4 x \langle 
{\rm tr}_{G/H}(\frac{1}{2} \mathscr{A}_\mu(x)^2)\rangle$.  
In light of these facts, our analyses in this paper is still insufficient, although  there is a claim that the Gribov copies inside the Gribov region have no influence on expectation-values of gauge invariant operators \cite{Zwanziger04}, in other words, the fundamental modular gauge fixing would be possible by the Langevin simulation, which does not seem to be confirmed e.g., by the lattice simulations. 
It is not clear whether the dependence on the gauge-fixing subset $H$ disappears.  Therefore, it is an open question whether the expectation value of the gauge dependent operator in the Landau gauge after fundamental modular gauge fixing, and that of the Maximal Abelian gauge should be the same after space-time averaging, i.e. sample averaging. 
The gauge-fixing independence could be checked by perturbation theory up to Gribov problem and by the  numerical simulations  where we need to remove the Gribov copies.

\section*{Acknowledgments}
The author would like to thank Yukinari Sumino and Kazuo Fujikawa for valuable comments on the relationship between gauge fixing and BRST invariance. 
This work is supported by 
Grant-in-Aid for Scientific Research (C)14540243 from Japan Society for the Promotion of Science (JSPS), 
and in part by Grant-in-Aid for Scientific Research on Priority Areas (B)13135203 from
the Ministry of Education, Culture, Sports, Science and Technology (MEXT).


\end{document}